\documentclass[showpacs,preprintnumbers,amsmath,amssymb, nofootinbib]{revtex4}

\usepackage{graphicx}
\usepackage{dcolumn}
\usepackage{bm}
\usepackage{color}

\usepackage{amsfonts,amsthm}

\newtheorem{teo}{Theorem}
\newtheorem{prop}[teo]{Proposition}

\def\al{\alpha}
\def\be{\beta}
\def\de{\delta}
\def\ga{\gamma}

\def\ep{\epsilon}

\def\te{\theta}
\def\la{\lambda}
\def\ze{\zeta}
\def\om{\omega}

\def\Ga{\Gamma}

\def\La{\Lambda}

\def\Si{\Sigma}

 \def\calC{{{\mathcal C}}}

 \def\one{{{\mathbb I}}}

 \def\D{{{\mathbb D}}}
 \def\C{{{\mathbb C}}}
 
 \def\E{{{\mathbb E}}}
 \def\H{{{\mathbb H}}}
 
 \def\F{{{\mathbb F}}}
 \def\G{{{\mathbb G}}}

 \def\K{{{\mathbb K}}}

\def\Aut{{\hbox{Aut}}}

\def\Spin{{\hbox{Spin}}}
\def\SO{{\hbox{SO}}}

\def\di{{\hbox{d}}}

\def\ip{\hbox to4pt{\leaders\hrule height0.3pt\hfill}\vbox to8pt{\leaders\vrule width0.3pt\vfill}\kern 2pt}
 
\def\del{\partial}
\def\na{\nabla}

\def\Lie{\pounds}

\def\arr{\rightarrow}

\def\then{\Rightarrow}

\def\ffrac[#1/#2]{\hbox{$\frac{#1}{#2}$}}

\def\({\left(}
\def\){\right)}
\def\[{\left[}
\def\]{\right]}
\def\^#1{{}^{#1}_{\>\cdot}}
\def\_#1{{}_{#1}^{\>\cdot}}
\def\<{\kern -1pt}
\def\z{e}

\begin{document}

\newpage

\title{Symmetry operators and separation of variables for Dirac's equation \\ on two-dimensional  spin manifolds with external fields}

\author{Lorenzo Fatibene}
 \email{lorenzo.fatibene@unito.it}
  \affiliation{Dipartimento di Matematica, Universit\`a di Torino, Italy\\
 INFN Sezione Torino- Iniz.~Spec.~Na12}

\author{Raymond G.\ McLenaghan}
 \email{rgmclena@uwaterloo.ca}
 \affiliation{Department of Applied Mathematics, University  of Waterloo, Waterloo, Ontario, N2L~3G1, Canada}
 
\author{Giovanni Rastelli}
 \email{giovanni.rastelli@unito.it}
 \affiliation{Dipartimento di Matematica, Universit\`a di Torino, Italy}


\begin{abstract}
The second order symmetry operators that commute with the Dirac operator with external vector, scalar and pseudo-scalar
potentials are computed on a general two-dimensional spin-manifold.  It is shown that the operator is defined in terms of
Killing vectors, valence two Killing tensors and scalar fields defined on the background manifold.  The commuting operator that
arises from a non-trivial Killing tensor is determined with respect to the associated system of Liouville  coordinates
and compared to the the second order operator that arises from that obtained from the unique separation scheme associated with
such operators.  It shown by the study of several examples that the operators arising from these two approaches coincide.
\end{abstract}


\maketitle

\def\FieldEqsEquivalenceAPP{A}
\def\FrameAPP{B}
\def \ProjectionIdentitiesAPP{C}

\section{Introduction}

The two-dimensional  Dirac equation is currently of great interest due to the connections with graphene's physics \cite{Nat2} and other experimental studies \cite{Nat1}.
It is   known  that the existence of  multiplicatively separated  solutions of the Dirac equation $\mathbb D\psi=\mu\psi$ in a given coordinate system and frame implies the additive separability in the same coordinates of the geodesic Hamilton-Jacobi equation. The logical chain of implications works as follows: if the Dirac equation admits multiplicatively separated solutions, then so does the squared Dirac equation; the highest-order terms of this equation coincide with those of the Laplace-Beltrami operator $\Delta$ acting on each component $\psi_i$ of the solution. Indeed, it can be shown \cite{M,S} that in this case  the Helmholtz  equation $\Delta \psi= \mu^2\psi$ must admit separable solutions. The multiplicative separation  of the last equation is possible only if the same coordinates allow the additive separation  of the geodesic Hamilton-Jacobi equation and the Ricci tensor is diagonalized in the same coordinates (Robertson condition)(\cite{BCR0} and references therein). There exists examples of separable Schr\"odinger equations whose corresponding Dirac equations are not separable \cite{Vi, Mc}.
Among the relevant steps towards a general theory we mention the works by Miller \cite{M} and Shapovalov and Ekle \cite{Sc} where a theory of complete separation associated with first-order symmetry operators of the Dirac operator is developed. 
Several contributions to the search for exact solutions of the Dirac equation in curved spaces may be found in Bagrov and Gitman \cite{Bag}  and  Cook \cite{Co}. In Shishkin  \cite{Sh, Sh3} and Shiskin et al. \cite{Sh0,Sh1,Sh2} an algebraic procedure is developed to obtain separation relations for the solution of the Dirac equation. The definition of separation used therein  coincides essentially with our definition of "naive separation" \cite{MR} in dimension two. It seems that any definition   of separation of variables in  two dimensions would yield the same results.
In \cite{KMW} it is proved that in Minkowski spaces all second-order symmetry operators of the Dirac operator (without external fields) can be factorized into products of the first-order operators.  It thus seemed that the problem of separation could be reduced to  separation associated only to first-order operators. Howewer, Fels and Kamran showed in  \cite{FK}  that this is not the case. Indeed, even if a second-order symmetry operator can be factorized into first-order ones,  these factors do not represent all the possible ways to separate the  Dirac equation, but only some of them.   
The existence of second-order symmetry operators is therefore important. The separation constants  that finally appear in the separated spinors solution  must be dynamical constants of the system, as also happens in the classical theory of separation for Hamilton-Jacobi and Schr\"odinger equations. It seems that the only way to obtain such dynamical constants is as  eigenvalues of symmetry operators of the Dirac operator, again in analogy with the separation of Schr\"odinger equation. In \cite{MR}  this ansatz is employed for the separation of the Dirac equation in  two-dimensional Riemannian manifolds. The separation relations are used to build first- and second-order symmetry operators; they arise  naturally from the decoupling of the separation relations. It turns out that not all the separation relations lead to symmetry operators. Some of these relations are connected only to second-order symmetry operators, in accordance with \cite{FK}. It  also follows that  at least one of the variables is  first-class, that is ignorable modulo rescalings. These results are  extended and refined in \cite{FMRS} and \cite{CFMR}, where a complete characterization of first and second-order symmetry operators of the Dirac operator is achieved in two-dimensional Riemannian and Lorentzian manifolds.

In the present paper we assume, as previously, that the Dirac equation is of eigenvalue-type, that is
$$
\mathbb D\psi =\mu \psi.
$$
Consequently, we consider separately the operator $\mathbb D$, where the eigenvalue $\mu$ does not appear, and the above equation. We make the same distinction between any second-order symmetry operator $\mathbb K$ and the  eigenvalue-type equation 
$$
\mathbb K \psi=\nu  \psi.
$$
The distinction is mathematically relevant because, otherwise, we characterize $\mathbb D$ and $\mathbb K$ as operators depending on $\mu$ and $\nu$ and consequently the eigenvalues are no longer free parameters labelling distinct solutions of the same Dirac equation but become parameters determining distinct Dirac equations, each one with a family of solutions depending on less parameters than the free case.  The separation of variables that we consider here is called complete in \cite{M,S} and depends on $nm$ free parameters, where $m$ is the dimension of the space of spinors, which is equal to equal to two in our case, and $n$ is the dimension of the configuration manifold. 
The separability property described above \cite{MR} is characterized invariantly   in terms of  second-order differential operators constructed from valence-two Killing tensors that commute  with the Dirac operator and admit the separable solutions as eigenfunctions with the separable constants as eigenvalues \cite{FMRS,CFMR}. In these works we obtain the most general second-order linear differential operator that commutes with the Dirac operator on a general two-dimensional pseudo-Riemannian manifold. Further it is shown that the operator is characterized in terms of Killing vectors and valence-two Killing tensors defined on the background manifold. The derivation is manifestly covariant: the calculations are done in general orthonormal frame independent of the choice of Dirac matrices.
 
The purpose of the present paper is to extend the results to the case when external vector, scalar and pseudo-scalar potentials are included.  

The paper is organized as follows: in Section II we summarize the basic properties of two-dimensional spin manifolds required for subsequent calculations.  Section III is devoted to the derivation of the form of the general second-order linear differential operator which commutes with the Dirac operator, we show that this operator is characterized by a valence two Killing tensor field, two Killing vector fields and two scalar fields defined on the background spin manifold. In Sections IV and V we determine the forms of the non-trivial first-order and second-order symmetry operators and we determine their integrability conditions in general coordinates.  The Killing tensor  characterizing second-order symmetry operators implies the existence of a system of canonical orthogonal coordinates called Liouville coordinates, in which the Killing tensor  is diagonal.  In Section VI we determine the coefficients of the non-trivial second-order symmetry operator in this system of coordinates by solving the determining equations and their integrability conditions. Section VII we apply to the present case the separability conditions determined in \cite{MR} for the uniqe separation scheme associated with second-order symmetry operators. Several examples are studied and it is shown that the second-order symmetry operator obtained agrees with that constructed in Section VI. The Conclusion is contained in Section VIII.

\section{Spin manifolds}

Let us consider a signature $\eta=(r,s)$ for dimension $m=r+s$.
Let us  denote by $\eta_{ab}$ the corresponding  canonical symmetric tensor.
 {By an abuse of language we shall also denote by $\eta$ the determinant of the bilinear form $\eta_{ab}$.}
A representation of the Clifford algebra $\calC(\eta)$ is induced by a set of Dirac matrices $\ga_a$ such that
\begin{equation}
\ga_a \ga_b + \ga_b \ga_a =2 \eta_{ab} \one
\label{DiracIdentitiesEQ}
\end{equation}
with $a,b, ...=0, \ldots, m-1$.

We stress that we shall not fix a particular set of Dirac matrices until Section {\bf VII} when we  consider schemes for  separation of variables. Until then we shall use only the algebraic consequences of (\ref{DiracIdentitiesEQ}).
The {\it even Clifford algebra} is spanned by the following matrices, namely  $\one$, $\gamma_a$, $\ga_{ab}:= \ga_{[a} \ga_{b]}$, \dots .  
The corresponding group $\Spin(\eta)$ is a multiplicative group in  $\calC(\eta)$.
One can define a covering map $\ell:\Spin(\eta)\arr \SO(\eta)$ by showing that for any element $S$ of $\Spin(\eta)$ one has $S\ga_a S^{-1}= \ell_a^b \ga_b$
with the matrix $\ell_a^b$ is  in $\SO(\eta)$.
We stress that until now we are at a purely algebraic level.

Let now $M$ be a connected, paracompact, $m$ dimensional spin manifold .
Let $P\arr M$ be a suitable $\Spin(\eta)$-principal bundle, such that it allows global (principal) morphisms $\z:P\arr L(M)$ of the {\it spin bundle} $P$ into the {\it general frame bundle} $L(M)$.
The local expression of such maps is given by {\it spin frames} $\z_a^\mu$; see \cite{FF}, \cite{FF1}. 
Let  $\z^a_\mu$ denote the inverse matrix of the spin frame $\z_a^\mu$.
A spin frame induces a metric $g_{\mu\nu}= \z^a_\mu\> \eta_{ab}\> \z^b_\nu$ and a spin connection
\begin{equation}
\Ga^{ab}_\mu = \z^a_\al \( \Ga^\al_{\be\mu} \z^{b\be} + \partial_\mu \z^{b\al}\)
\end{equation}
where $ \Ga^\al_{\be\mu}$ denotes the Levi-Civita connection of the induced metric $g_{\mu\nu}$.
We note that such a connection satisfies $\na_\mu \z_a^\nu=\del_\mu \z_a^\nu + \Ga^\nu_{\la\mu} \z_a^\la + \Ga^b{}_{a\mu}\z^\nu_b=0$,
and  is antisymmetric in the upper indices $[ab]$.
We also remark that Latin indices are raised and lowered by the inner product $\eta_{ab}$ while Greek indices are raised and lowered by the induced metric 
$g_{\mu\nu}$.
For subsequent use we introduce the frame covariant derivative $\na_a:=\z_a^\mu \na_\mu$. 

If an electromagnetic field is allowed then a covariant potential $A_\mu(x)$ is to be considered.
The electromagnetic field $A_\mu$ is a principal connection on a suitable $U(1)$-bundle $Q$.
In this setting the spinor fields $\psi$ are sections of a bundle $\Si$ associated to $P\times_M Q$.
The field strength of the electromagnetic field will be defined as
 \begin{equation}
F_{\mu\nu}=\del_\mu A_\nu-\del_\nu A_\mu
\qquad\qquad
F_{ab}= \z_a^\mu \z_b^\nu F_{\mu\nu}
\end{equation}

If a (matrix) potential is allowed it will be denoted by a function ${\bf V}(x)$.
Of course, the mass is a particular case of scalar potential and can be merged into the potential function ${\bf V} = m^2\one$.

The Dirac equation then has the form
\begin{equation}
\D\psi= i \ga^a D_a \psi -{\bf V}(x)\psi =0,
\label{DiracEquationEQ}
\end{equation}
where the gauge covariant derivative of the spinor $\psi$ is defined as 
\begin{equation}
D_\mu \psi= \del_\mu \psi +\ffrac[1/4]  \Ga^{ab}_\mu \ga_{ab}\> \psi -iq A_\mu \psi
\qquad\qquad
D_a:=\z_a^\mu D_\mu
\label{CovDer}
\end{equation}


A {\it gauge transformation} is an automorphism of the bundle $P\times_M Q$, i.e.~locally
\begin{equation}
\begin{cases}
 x'= f(x) \\
 g'= S(x)\cdot g\\
 e^{i\te'}= e^{i\al(x)}\cdot e^{i\te}\\
\end{cases}
\end{equation}
with $e^{i\al(x)}\in U(1)$ and $S(x)\in \Spin(\eta)$.

Gauge transformations form a group denoted by $\Aut(P\times_MQ)$ which acts on spinors, frame, spin connection and electromagnetic field by
\begin{equation}
\begin{cases}
\psi'= e^{i\al}\>S\cdot\psi \\
\z'{}_a^\mu= J^\mu_\nu \z_b^\nu \ell_a^b(S)
\qquad\then\Ga'{}^{ab}_\mu= \bar J_\mu^\nu \ell^a_c(S)\( \Ga^{cd}_\nu \ell^b_d(S) + \di_\nu \ell^c_d(S)\> \eta^{db}\)\\
A'_\mu= \bar J_\mu^\nu \( A_\nu + \del_\nu \al \)\\
\end{cases}
\end{equation}
leaving the Dirac equation (\ref{DiracEquationEQ}) invariant. 
Here $J^\mu_\nu$ is the Jacobian matrix of the spacetime transfomation $x'= f(x)$ onto which the spin transformation projects and $\bar J^\mu_\nu$ denotes the anti-Jacobian. The covering map is denoted by $\ell$ as above.
The  potential ${\bf V}$ transforms under gauge transformations as a scalar field.


Depending on the object on which it acts, the covariant derivative $D_\mu$ may or may not depend on the electromagnetic potential $A_\mu$.
If the object on which the covariant derivative acts is insensitive to phase gauge shifts (for example as spinors if one set the charge $q$ to zero)
then we shall denote the covariant derivative as $D_\mu=\na_\mu$. In other words $\na_\mu$ will be used to emphasize that covariant derivative does not 
depend on $A_\mu$. For example $U(1)$-gauge transformations have no effect on the spin frame $\z_a^\nu$; accordingly, the
covariant derivative of the frame is denoted by $\na_\mu \z_a^\nu$.

For future notational convenience we can also set
\begin{equation}
\na_\mu \psi= \del_\mu \psi +\ffrac[1/4]  \Ga^{ab}_\mu \ga_{ab}\> \psi
\end{equation}
so that one has $D_\mu \psi=\na_\mu \psi-iq A_\mu \psi$. 
However, this is an abuse of language, since $\na_\mu \psi$ does not transform properly under $U(1)$-gauge transformations (unless $q=0$).
On the other hand, purely spacetime diffeomorphisms (with $U(1)$-gauge transformations set to the identity) are not global transformations unless the bundle $Q$ is trivial (and, in any case, they are gauge dependent) .


Let us show that the Dirac matrices $\ga^a$ are left invariant by spin transformations.
For, one has
\begin{equation}
D_\mu \ga^a\equiv \na_\mu \ga^a= \del_\mu \ga^a + \Ga^a{}_{b\mu} \ga^b  + \ffrac[1/4]  \Ga^{cd}_\mu \ga_{cd}\> \ga^a - \ffrac[1/4]  \Ga^{cd}_\mu \ga^a \ga_{cd}=
 \Ga^a{}_{b\mu} \ga^b  -  \Ga^{ac}_\mu \ga_{c}\equiv 0
\end{equation}
where we used the identity $[\ga_{cd}, \ga^a]= 4\de^a_{[d} \ga_{c]}$.

The Dirac equation is left invariant also by change of bases in the space of spinors, namely
\begin{equation}
\begin{cases}
\psi'= P\cdot\psi \\
\ga'^a= P\cdot \ga^a P^{-1}\\
\end{cases}
\end{equation}
for any constant invertible matrix $P$. If $\ga_a$ are Dirac matrices then $\ga'_a$ are Dirac matrices as well.

A {\it second order symmetry operator} for the Dirac equation is an operator of the form
\begin{equation}
\K= \E^{ab}D_{ab} + \F^a D_a +\G \one 
\label{29}\end{equation}
which commutes with the Dirac operator $\D$.
Here $D_{ab} = \ffrac[1/2]\( D_a D_b + D_b D_a\)$ denotes the {\it symmetrized} second covariant derivative (expressed in the frame).
The coefficients $ \E^{ab},  \F^a, \G$ are matrix zero-order operators.
By expanding the condition $[\K, \D]=0$ one obtains
\begin{equation}
\begin{cases}
\E^{(ab}\ga^{c)}- \ga^{(c}\E^{ab)}=0 \\
\F^{(a}\ga^{b)}- \ga^{(b}\F^{a)}= \ga^c \na_c \E^{ab}-i {\(\E^{ab} {\bf V}- {\bf V}\E^{ab}\)} \\
\G\ga^{a}- \ga^{a}\G= \ga^c \na_c \F^{a} -i  {\(\F^{a} {\bf V}- {\bf V}\F^{a}\)}
 -\ffrac[1/4]\(\E^{ab}\ga^c + \ga^c \E^{ab}\)\ga_{ef} R^{ef}{}_{bc}
+\ffrac[1/3]\(\E^{ef}\ga^c - 2i\ga^c\E^{ef}\)R^a{}_{efc}+\\
\qquad\qquad
+iq\(\E^{ab}\ga^c + \ga^c \E^{ab}\) F_{bc} 
- {2i\E^{ab}\na_b {\bf V}}\\
\ga^c\na_c \G=i  {\(\G {\bf V}- {\bf V}\G\)}+
\ffrac[1/12]\na_a R^{ef}{}_{bc}\(2\E^{ab}\ga^c + \ga^c \E^{ab}\)\ga_{ef}
+\ffrac[1/8]\(\F^{a}\ga^b + \ga^b \F^{a}\)\ga_{ef} R^{ef}{}_{ab}+\\
\qquad\qquad
-\ffrac[iq/3]\(2\E^{ab}\ga^c + \ga^c \E^{ab}\)\na_a F_{bc}
-\ffrac[iq/2]\(\F^{a}\ga^b + \ga^b \F^{a}\) F_{ab}
+ {i \E^{ab}D_{ab} {\bf V}} + {i \F^a \na_a {\bf V}},
\\
\end{cases}
\label{Cond}
\end{equation}
Where we have used the results of  Appendix A.
The conditions (\ref{Cond}) do not depend on the dimension $m$ or on the signature $\eta$.
To obtain the form of the second order symmetry operator one has first to fix the dimension, use the Clifford identities,  what is known from differential geometry about the curvature tensors, and solve these equations.

\section{Second order symmetry operators in dimension 2}

In dimension $m=2$ the Clifford algebra is spanned by the elements $\one$, $\ga_a$, $\ga\ga_0\ga_1$. 
The Dirac matrices are linearly independent and so they provide a basis for $2\times 2$ $\C$-matrices.
Thus any matrix can be written as a linear combination of these matrices. 
Thus we may write
\begin{equation}\label{SO}
\begin{cases}
\E^{ab}:= e^{ab}\one + e^{ab}_c \ga^c + \hat e^{ab}\ga\\
\F^{a}:= f^{a}\one + f^{a}_c \ga^c + \hat f^{a}\ga\\
\G:= g\one + g_c \ga^c + \hat g\ga\\
\end{cases}
\end{equation}
where the coefficients are   tensors fields on the spin manifold. 
Also the matrix potential can be expanded in the basis as
\begin{equation}
 {{\bf V}= V\one + V_a \ga^a + \hat V \ga}
\end{equation}


The coefficient $V$ is called a {\it scalar potential}, $V_a$   a {\it vector potential} and $\hat V$  a {\it pseudopotential}.
The vector potential contributes to the Dirac operator for a term formally analogous to the electromagnetic field. Thus we can set it to zero without loss of generality.


In other words if $A_\mu$ is a good potential for an electromagnetic field and $V_a$ is a vector potential then one can define a new potential $q\tilde A_\mu= qA_\mu + \z_\mu^a V_a$ (for a different electromagnetic field) and neglecting the vector potential.
Accordingly, the vector potential can be neglected without loss in generality.


Moreover, in dimension $m=2$ the Riemann tensor has  the form
\begin{equation}
 R^{ab}{}_{cd}= \ffrac[1/2] R \ep^{ab}\ep_{cd}
\end{equation}
 where $R$ is the scalar curvature and $\ep_{cd}$ denotes the Levi-Civita tensor.
 Similarly, the field strength of the electromagnetic field may be written
 \begin{equation}
F_{cd}= F \ep_{cd}
\end{equation}
for some function $F:= F_{01}$.

Any product of Dirac matrices is a $2\times 2$ matrix and can hence be expanded as a linear combination of  $\one$, $\ga_a$, $\ga$. 
We  collect in Appendix B a number of useful formulae for manipulate products of Dirac matrices.

The first equation of (\ref{Cond}) yields
\begin{equation}
\begin{aligned}
 \E^{(ab}\ga^{c)}- \ga^{(c}\E^{ab)}=& \ga^d \ga^{(c} e^{ab)}_d- e^{(ab}_d  \ga^{c)}\ga^d
+ \ga \ga^{(c} \hat e^{ab)}- \hat e^{(ab}  \ga^{c)}\ga= \\
 =&-2 e^{(ab}_d \ep\^{c)}\^d \ga -2\eta \ep\^{(c}{}_d \hat e^{ab)} \ga^d=0\\
 \end{aligned}
\end{equation}

Since $\ga^d$ and $\ga$ are independent, this implies
\begin{equation}
\begin{cases}
 e^{(ab}_d \ep^{c)d}  =0\\
 \hat e^{(ab} \ep^{c)d} =0\\
 \end{cases}
 \label{FirstEq}
\end{equation}

The first condition in (\ref{FirstEq}) implies 
\begin{equation}
 e^{ab}_d  =2 \al^{(a} \de^{b)}_d
 \label{egammaa}
\end{equation}
 where $\al^a$ is some vector. 
In fact, by expanding the  first condition in (\ref{FirstEq}) one finds
\begin{equation}
\begin{aligned}
&\[e^{ab}_d \ep^{cd} +e^{bc}_d \ep^{ad} + e^{ca}_d \ep^{bd} =0\] (\cdot \ep_{cj})\\
&3e^{ab}_j   = 2 e^{d(b}_d \de^{a)}_j    \\
 \end{aligned}
\end{equation}
From which (\ref{egammaa}) follows by setting $\al^a= \ffrac[1/3]e^{ad}_d$.

The second condition in (\ref{FirstEq}) implies 
\begin{equation}
 \hat e^{ab}  =0
\end{equation}

In fact, by expanding the  second condition in (\ref{FirstEq}) one has
\begin{equation}
\begin{aligned}
&\[\hat e^{(ab} \ep^{c)d}  =0\] (\cdot \ep_{dj})
\qquad\then  \hat e^{(ab} \de^{c)}_j  =0 
\qquad\then 
\[ \hat e^{ab} \de^{c}_j + \hat e^{bc} \de^{a}_j+ \hat e^{ca} \de^{b}_j  =0 \] (\cdot \de^j_c)\\
&\qquad\then  4\hat e^{ab}  =0\\
 \end{aligned}
\end{equation}

Then, as a consequence of first equation of (\ref{Cond}), we have
\begin{equation}
\E^{ab}= e^{ab}\one + 2 \al^{(a} \ga^{b)} \\
\label{E1}
\end{equation}
for some vector $\al^a$.
As expected, until this point the solution is completely unaffected by the electromagnetic field and the scalar potentials and reduces to the result found previously in \cite{CFMR}.

Let us now focus on the second equation in (\ref{Cond}). By expanding the left hand side one obtains
\begin{equation}
\begin{aligned}
&\F^{(a}\ga^{b)}- \ga^{(b}\F^{a)}
= -2\eta f^{(a}_c \ep^{b)c}\ga -2 \hat f^{(a}\ep^{b)}{}\_c\ga^c \\
 \end{aligned}
\end{equation}

Similarly the first term on the right hand side yields
\begin{equation}
\begin{aligned}
& \ga^c \na_c \E^{ab}
= \na_c e^{ab} \ga^c + 2 \na^{(a}\al^{b)}\one -2\eta\na_c\al^{(a} \ep^{b) c}\ga\\
 \end{aligned}
\end{equation}

while  the second term gives
\begin{equation}
-i {\(\E^{ab} {\bf V}- {\bf V}\E^{ab}\)}=
 {-2i \(\al^{(a}\ga^{b)}\ga- \al^{(a}\ga\ga^{b)}\)\hat V}= {-4i \al^{(a} \ep^{b)c}\hat V \ga_c}
\end{equation}

Thus the second equation in (\ref{Cond}) is  equivalent to the conditions
\begin{equation}
\begin{cases}
  \na^{(a}\al^{b)}=0\\
  -2 \hat f^{(a}\ep^{b)c}=\na^c e^{ab}- {4i \al^{(a} \ep^{b)c}\hat V}
  \qquad\then
\begin{cases}
  \na^{(c} e^{ab)}=0\\
   \hat f^{a}\de^{b}_d + \hat f^{b}\de^{a}_d=\ep_{cd}\na^c e^{ab} - {2i\(\al^a\de^b_d+\al^b\de^a_d\)\hat V}\\
 \end{cases}\\
  f^{(a}_c \ep^{b)c}=\na_c\al^{(a} \ep^{b) c} \\
 \end{cases}
 \label{SecondEq}
\end{equation}
The first condition $ \na^{(a}\al^{b)}=0$ implies that the vector $\al^a$ is a Killing vector.

The second condition in (\ref{SecondEq}) 
implies that $e^{ab}$ is a Killling tensor. Moreover, by contracting by$\ep_{cb}$ one obtains
 \begin{equation}
 \hat f^{a} =\ffrac[1/3]\ep_{cb}\na^c e^{ab}+ {2i\al^a\hat V}
\end{equation}
which determines the coefficient $ \hat f^{a}$ in terms of the coefficient $e^{ab}$.

The third condition in (\ref{SecondEq}) may be expanded to yield
\begin{equation}
\begin{aligned}
& \[f^{a}_c \ep^{bc}+  f^{b}_c \ep^{ac}=\na_c\al^{a} \ep^{bc}+ \na_c\al^{b} \ep^{ac} \] (\cdot \ep_{bd})\\
& 2f^{a}_d  - f^{b}_b \de^a_d =2\na_d\al^{a}-\na_c\al^{c}\de^a_d \\
& f^{a}_c   =  \al \de^a_c+ \na_c\al^{a}  \\
 \end{aligned}
\end{equation}
where we set $\al:= \ffrac[1/2]f^{b}_b$.

Thus, as a consequence of first and second equations of (\ref{Cond}), we have
\begin{equation}
\begin{cases}
\E^{ab}= e^{ab}\one + 2 \al^{(a} \ga^{b)} \\
\F^{a}:= f^{a}\one + \( \al  \de^a_c+ \na_c\al^{a} \) \ga^c + \(\ffrac[1/3]\ep_{cb}\na^c e^{ab} + {2i\al^a\hat V}\)\ga\\
\end{cases}
\qquad\qquad 
\begin{aligned}
& \na^{(c} e^{ab)}=0\\
&  \na^{(a}\al^{b)}=0\\
& \al: M\arr \C
 \end{aligned}
\end{equation}
When no electromagnetic field and   potentials are present we recover the results of  \cite{CFMR}.

The third equation in (\ref{Cond}) is equivalent to the conditions
\begin{equation}
\begin{cases}
  \na_a \al  = \om_a\\
  2g_c \ep^{ac}=\(\na_c \al -\ffrac[1/2] R\al_c\) \ep^{ac}  + {2i\eta e^{ab} \na_b\hat V}\\
  2 \hat g\ep^{ca} =\na^c f^a -\ffrac[1/3] \( \na_b\na^c e^{ab}- \na_b\na^b e^{ac} \) + R e^{ac} - \ffrac[1/2] R e \eta^{ac}
  -2iq F \ep\^c{}_b e^{ab} -  2i( \al^{a}  \na^c V   + \al^{b}  \na_b V  \eta^{ac}) +\\
  \qquad\quad\>\>
   +  { 2i \(\na_b (\al^a \hat V) \ep^{bc} -\al^{a} \na_b \hat V\ep^{bc}-\al^{b} \na_b \hat V\ep^{ac}\)  -2i\(\al\ep^{ac} + \na_b\al^a \ep^{bc}\)\hat V}
 \end{cases}
 \label{ThirdEq}
\end{equation}
where we set $e:=  e^{ab} \eta_{ba}$ and
\begin{equation}
\om_a:=2i q F\ep_{ab}\al^b + 2i e\_a{}^{b} \na_b V
\end{equation}
When electromagnetic field and   potentials vanish then the first of these conditions reduces to $\al\in \C$ as previously found in  \cite{CFMR}.

In general, this condition can be written as $d \al=\om_a \z^a:= 2i(q F \ep_{ab}\al^b +  e\_a{}^{b} \na_b V) \z^a$
which has a (local) solution iff the integrability condition $d\om=0$ is satisfied. 
This integrability condition reads as 
 \begin{equation}
\na_{[d} \om_{a]}=2i\[ q  \ep_{b[a}\na_{d]} (F\al^b)+  \na_{[d}( e\_{a]}{}^{b} \na_b V)\]=0
\qquad\then
 q \al^b\na_{b} F=-  \ep^{da}\na_{d}( e\_{a}{}^{b} \na_b V)
\end{equation}
This is a condition on the electromagnetic field and scalar potentials for the existence of second order symmetry operator.
If  $\al^b$ or $e^{ab}$  vanish the integrability condition simplifies. 
We shall postpone the analysis of the integrability conditions until Section IV.

The second condition in (\ref{ThirdEq}) may be written as
\begin{equation}
g_a=\ffrac[1/2]\na_a \al -\ffrac[1/4] R \al_a  - {i\eta e^{cb} \na_b\hat V \ep_{ac}} 
=iq F \ep_{ab}\al^b +  ie\_a{}^{b} \na_b V -\ffrac[1/4] R \al_a  - {i\eta e^{cb} \na_b\hat V\ep_{ac}}
\end{equation}
which determines the coefficient $g_a$.

We split  the third condition in  (\ref{ThirdEq}) into symmetric and antisymmetric parts to obtain
\begin{equation}
\begin{cases}
  0 =\na^{(c} f^{a)} -\ffrac[1/3] \( \na_b\na^{(c} e^{a)b}- \na_b\na^b e^{ac} \) + R e^{ac} - \ffrac[1/2] R e \eta^{ac}
  -2iq F \ep\^{(c}{}_b e^{a)b} -  2i( \al^{(a}  \na^{c)} V   + \al^{b}  \na_b V  \eta^{ac}) \\
4 \hat g =\na^c f^a \ep_{ca}-\ffrac[1/3]  \na_b\na^c e^{ab} \ep_{ca} 
  -2iq F \ep\^c{}_b e^{ab} \ep_{ca} -  2i \al^{a}  \na^{c} V \ep_{ca}  
   +  { 4i \al^{b} \na_b \hat V +4i\al\hat V}\\
 \end{cases}
 \label{eq31}
\end{equation}

By using identity (\ref{L1}) in the second of these conditions,  one finds that
\begin{equation}
 \hat g=\ffrac[1/4](\na^c \ze^a  -2iq F \ep\^c{}_b e^{ab} -  2i \al^{a}  \na^c V   )\ep_{ca} +  { i  \al^a \na_a \hat V + i\al \hat V }
\end{equation}
where we set $\ze^a:=f^{a} -\na_b e^{ab}$.
The first condition in (\ref{eq31}) can be recasted as
\begin{equation}
\na^{(c} \ze^{a)} =
  2iq F \ep\^{(c}{}_b e^{a)b} +  2i( \al^{(a}  \na^{c)} V   + \al^{b}  \na_b V  \eta^{ac})  =:\ffrac[1/2] \La^{ca}
  \label{zKilling}
\end{equation}
Notice that $\La:=\La^{ca} \eta_{ca}=  4iq F \ep_{ab} e^{ab}+  12 i\al^{a}  \na_{a} V   = 12 i\al^{a}  \na_{a} V$.

When there is no electromagnetic field or  potentials, the equation (\ref{zKilling})  reduces to $\na^{(c} \ze^{a)} =0$ which implies that $\ze$
is a Killing vector, as found in  \cite{CFMR}. 
In this more general situation we need to discuss whether it is possible to solve   (\ref{zKilling})  for  $\ze^a$.

We now consider this issue in greater detail. First we notice that   (\ref{zKilling}) can be recast in the form
$\Lie_\ze g_{\al\be}= \La_{\al\be}$ and that in general one has
\begin{equation}
 \Lie_\ze\{g\}^\al_{\be \mu}= \ffrac[1/2]g^{\al\la}\(-\na_\la \Lie_\ze g_{\be\mu}+\na_\be \Lie_\ze g_{\mu\la} + \na_\mu \Lie_\ze g_{\la\be}\)
 = \ffrac[1/2]g^{\al\la}\(-\na_\la \La_{\be\mu}+\na_\be \La_{\mu\la} + \na_\mu \La_{\la\be}\)
\end{equation}
where $\{g\}^\al_{\be \mu}$ denote the Christoffell symbols of the metric $g_{\mu\nu}$.
Then we have 
\begin{equation}
 \Lie_\ze R^\al{}_{\be \mu\nu}= \na_\mu \Lie_\ze\{g\}^\al_{\be \nu}- \na_\nu \Lie_\ze\{g\}^\al_{\be \mu}
 = \ffrac[1/2] \Lie_\ze R \ep_{\mu\nu}\ep^\al\_\be + {\ffrac[R/2] \ep_{\mu\nu} \ep^{\al\la}\Lie_\ze g_{\la\be}}
\end{equation}
from which one obtains
\begin{equation}
 \Lie_\ze R =
 \na_c \na^b \La_d\^a{}  \ep^{cd}  \ep_{ab} 
 -  {\ffrac[R/2]  \La} 
 \label{39}
\end{equation}
Since $R$ is a scalar, so that $\Lie_\ze R =\ze^a\na_a R$,  and under the assumption that   (\ref{zKilling}) has a solution then it follows that
\begin{equation}
 \ze^a\na_a R =\na_c \(\na^b \La\_{d}{}^a \)  \ep^{cd}\ep_{ab} - {\ffrac[R/2]\La}
\end{equation}
which is an integrability condition for equation (\ref{zKilling}). For example, on constant curvature spaces one finds that
\begin{equation}
\na_c \(\na^b \La\_{d}{}^a \)  \ep^{cd}\ep_{ab}= {\ffrac[R/2]\La}
\end{equation}
is a necessary condition for   (\ref{zKilling})  to have a solution. We shall continue the study of  these integrability conditions in Section VI. 

We now  summarize  what we have found.
As a consequence of first, second and third equations of (\ref{Cond}), we have that 
\begin{equation}
\begin{cases}
\E^{ab}= e^{ab}\one + 2 \al^{(a} \ga^{b)} \\
\F^{a}:= (\ze^{a} +\na_b e^{ab})\one + ( \al \de^a_c+ \na_c\al^{a}) \ga^c +\(\ffrac[1/3]\ep_{cb}\na^c e^{ab} + {2i\al^a\hat V}\)\ga\\
\G:= g\one + \(iq F \ep_{ab}\al^b +  ie\_a{}^{b} \na_b V -\ffrac[1/4] R \al_a  + {i\eta e^{cb} \na_b\hat V\ep_{ac}}\) \ga^a +\\
\qquad+ \(\ffrac[1/4]\(\na^c \ze^a  -2iq F \ep\^c{}_b e^{ab} -  2i \al^{a}  \na^c V   \)\ep_{ca} +  { i  \al^a \na_a \hat V +i\al \hat V}\)\ga\\
\end{cases}
\end{equation}
where the coefficients satisfy the conditions
\begin{equation}
\begin{aligned}
& \na^{(c} e^{ab)}=0\\
&  \na^{(a}\al^{b)}=0\\
 \end{aligned}
 \qquad\qquad\qquad
\begin{aligned}
& \na_a \al=\om_a\\
& \na^{(c} \ze^{a)} =\ffrac[1/2] \La^{ca}
 \end{aligned}
\end{equation}
where 
\begin{equation}
\begin{cases}
\La^{ca}=  4i(q F \ep\^{(c}{}_b e^{a)b} +   \al^{(a}  \na^{c)} V   + \al^{b}  \na_b V  \eta^{ac})\\
\om_a= 2i \(q F \ep_{ab}\al^b + e\_a{}^{b} \na_b V\)\\
\end{cases}
\end{equation}
and  the following integrability conditions are satisfied
\begin{equation}
\begin{cases}
\na_{[a}\om_{b]}=0
\qquad\qquad\qquad\qquad\then
q \al^b\na_{b} F=-  \ep^{da}\na_{d}( e\_{a}{}^{b} \na_b V)
\\
 \ze^a\na_a R =\na_c \(\na^b \La^{da} \)  \ep^{c}\_d\ep_{ab} - {\ffrac[R/2]\La}\\
 \end{cases}
\end{equation}

Finally we consider the fourth and last equation in (\ref{Cond}).
It is equivalent to the conditions
\begin{equation}
\begin{cases}
\ze^{a}\na_a V = {-\eta\(\ffrac[2/3] \na_a e^{bc}\na_b\hat V \ep\^a{}_b +e^{bc}\na_{ac}\hat V\ep\^a{}_b - i\al^a\na_a(\hat V)^2\)}\\
 {\ze^a\na_a\hat V}=\ffrac[2/3]\ep_{cb}\na^c e^{ab}  \na_a V+ \ep\^c{}_{b} e^{ab}  \na_{ca} V
    +  {\na_c(e^{cb}\na_b\hat V)  -2i\al^a\na V \hat V} \\
 \na_c g' =
  i q F\ep_{ca} \ze^a -\ffrac[1/4]\na_a\(Re^a\_c\)  -i\om_c  V+ {i\al^a\na_{ab}\hat V\ep^b{}\_c -\ffrac[i/2] R\al_a \hat V \ep^a{}\_c +\eta e\_c{}^b\na_b(\hat V)^2} =: \La_c
 \end{cases}
 \label{FourthEq}
\end{equation}
where we set $g' :=g - 3i\al^a \na_a V - i\al  V$.

Thus (\ref{29}) is a second order symmetry operator iff
\begin{equation}
\begin{cases}
\E^{ab}= e^{ab}\one + 2 \al^{(a} \ga^{b)} \\
\F^{a}:= (\ze^{a} +\na_b e^{ab})\one + ( \al \de^a_c+ \na_c\al^{a}) \ga^c +\(\ffrac[1/3]\ep_{cb}\na^c e^{ab} + {2i\al^a\hat V}\)\ga\\
\G:= (g' + 3i\al^a \na_a V +  i\al  V)\one + \(iq F \ep_{ab}\al^b +  ie\_a{}^{b} \na_b V -\ffrac[1/4] R \al_a  - {i\eta e^{cb} \na_b\hat V\ep_{ac}}\) \ga^a +\\
\qquad+ \(\ffrac[1/4]\(\na^c \ze^a  -2iq F \ep\^c{}_b e^{ab} -  2i \al^{a}  \na^c V   \)\ep_{ca} +  { i  \al^a \na_a \hat V +i\al \hat V}\)\ga\\
\end{cases}
\qquad
\begin{aligned}
& \na^{(c} e^{ab)}=0\\
&  \na^{(a}\al^{b)}=0\\
&\na_a \al=\om_a\\
& \na^{(c} \ze^{a)} =\ffrac[1/2] \La^{ca}\\
& \na_a g'= \La_a
 \end{aligned}
 \label{SOSOP}
\end{equation}
where we set
\begin{equation}
\begin{cases}
\La^{ca}=  4i\(q F \ep\^{(c}{}_b e^{a)b} +   \al^{(a}  \na^{c)} V   + \al^{b}  \na_b V  \eta^{ac}\)\\
\La_c=  i q F\ep_{ca} \ze^a -\ffrac[1/4]\na_a\(Re^a\_c\)  -i\om_c  V+ {i\al^a\na_{ab}\hat V\ep^b{}\_c -\ffrac[i/2] R\al_a \hat V \ep^a{}\_c +\eta e\_c{}^b\na_b(\hat V)^2} \\
 \om_c= 2i(q F \ep_{cd}\al^d +  e\_c{}^{d} \na_d V)\\
\end{cases}
\end{equation}
and  the following integrability conditions are satisfied
\begin{equation}\label{49}
\begin{cases}
\ep^{bc }\na_{b}\om_{c}=0
\qquad\then
q  \al^c\na_{c}F = -\ep^{ac}\na_{c}( e\_a{}^{b} \na_b V)\\
\ep^{dc} \na_{d} \La_{c}=0
\\
 \ze^a\na_a R =\na_c \(\na^b \La^{da} \)  \ep^{c}\_d\ep_{ab}  - {\ffrac[R/2]\La}\\
\ze^{a}\na_a V =- {\eta\(\ffrac[2/3] \na_a e^{bc}\na_c\hat V \ep\^a{}_b +e^{bc}\na_{ac}\hat V\ep\^a{}_b - i\al^a\na_a(\hat V)^2\)}\\
 {\ze^a\na_a\hat V}=\ffrac[2/3]\ep_{cb}\na^c e^{ab}  \na_a V+\ep\^c{}_{b}\ e^{ab} \na_{ca} V   -  { 2i\al^a\na_a V \hat V} \\
 \end{cases}
\end{equation}

\section{First order operators}

Before discussing the existence of second order symmetry operators, we  discuss first order (and zero order) symmetry operators.
First order symmetry operators are obtained by setting above $\al^a=0$ and $e^{ab}=0$.
For first order operators $\hat \K^{(1)}$ one obtains
\begin{equation}
\begin{cases}
\F^{a}:= \xi^{a}\one + a \ga^a \\
\G:= (\om +  i a  V)\one +\( \ffrac[1/4]\na^c \xi^a   \ep_{ca}+ {ia\hat V}\)\ga\\
\end{cases}
\qquad
\begin{aligned}
& a\in\C\\
& \na^{(c} \xi^{a)} =0\\
& \na_a \om= \La_a
 \end{aligned}
\end{equation}
where we set $\La_c:=iq F\ep_{cb} \xi^b$ and we have the following integrability conditions 
\begin{equation}
\begin{cases}
 \xi^a\na_a R =0
 \qquad\Leftarrow \na^{(c} \xi^{a)} =0\\
\ep^{ac} \na_{a} \La_{c}=0
 \qquad\iff
\xi^a \na_a F =0 \\
 \xi^{a}\na_a V  =0\\
 { \xi^{a}\na_a \hat V  =0}\\
 \end{cases}
\end{equation}

For zero order operators one obtains $\K^{(0)}:= k\one$, with $k\in \C$, and no integrability conditions.

Among first order operators one has the product of a zero order operator with the Dirac operator, i.e.~$\H^{(1)}:=k \D= ik \ga^a\na_a - k {\bf V}= ik \ga^a\na_a - kV -   {k\hat V\ga}$, which are trivially symmetry operators
and they correspond to the choice $\F^a=ik \ga^a$, $\G=-kV\one - { k\hat V \ga}$, i.e.~to the choice $\xi^a=0$, $a=ik$, $\om=0$.

Non-trivial first order symmetry operators  then have the form
\begin{equation}\label{fokv}
\K^{(1)}=  \(\xi^{a} \na_a + \om \)\one + \ffrac[1/4]\na^c \xi^a   \ep_{ca}\ga
\end{equation}
where $\xi^a$ is a (non-zero) Killing vector, $\om$ is a function such that $\na_c \om= -iq F\ep_{ac} \xi^a$ and the following integrability conditions are satisfied:
\begin{equation}\label{iceq1}
\begin{cases}
 \xi^b\na_b F=0 \\
 \xi^{a}\na_a V  =0\\
 { \xi^{a}\na_a \hat V  =0}\\
 \end{cases}
\end{equation}
since, in view of (\ref{39}), the condition $\xi^a\na_a R =0$ is a consequence of $\na^{(c} \xi^{a)} =0$.
The integrability conditions select the electromagnetic field and the scalar potentials which are compatible with existence of non-trivial first order symmetry operators.
The function $\om$ satisfying $\na_c \om= iq F\ep_{ca} \xi^a$ always exists (in view of $\xi^b\na_b F=0 $) and it is defined modulo a constant which corresponds to a zero order symmetry operator $\K^{(0)}$.

Thus one has the following:

\medskip

\noindent
{\bf Theorem: }
The existence os  a non-trivial first order symmetry operator $\K^{(1)}$ implies that $M$ admits a non-zero Killing vector $\xi^a$.

\medskip

Analogously, for any first order symmetry operator $\hat\K^{(1)}$ one can define a trivial second order symmetry operator $\H^{(2)}= \hat \K^{(1)}\> \D$. 
This is in the form:
 \begin{equation}
 \begin{aligned}
 \H^{(2)}=& \( (\xi^{a}\one + a \ga^a)\na_a +((\om -  ia  V)\one +\( \ffrac[1/4]\na^c \xi^a   \ep_{ca} + {ia\hat V}\) \ga)\)\(i\ga^b\na_b - V\one - {\hat V\ga}\)=  \\
=& i(a\eta^{ab} + \xi^{(a} \ga^{b)})\na_{ab}
+\(\(-\xi^aV\) \one + \((i\om-2aV)\de^a_c +\ffrac[i/2]\na_c\xi^a\)\ga^c - {\(\xi^a\hat V\)\ga}\)\na_a+\\
&+(-\ffrac[ia/4]R + {\ffrac[1/4]\eta\na^k\xi^a\ep_{ka}\hat V} -\om V -iaV^2 +  {i \eta a \hat V^2} )\one
+(\ffrac[q/2]\xi^a F \ep_{ak} -\ffrac[i/8]R\xi_k -a\na_k V - {a\na_b \hat V\ep^b\_k})\ga^k+\\
&+(\eta a q F - \ffrac[1/4]\na^c\xi^a\ep_{ca} V- {\om \hat V} - {2iaV\hat V})\ga\\
 \end{aligned}
\end{equation}
This operator corresponds to specifying a general second order operator (\ref{SOSOP}) by setting
$e^{ab}=ia\eta^{ab}$, $\al^a=\ffrac[i/2]\xi^a$, $\ze^a= - V\xi^{a}$, $\al = i \om-2aV$, $g'=-\ffrac[ia/4]R+ia V^2
+ {i\eta a\hat V^2} - {\ffrac[1/4]\eta \na^k\xi^a\ep_{ka}\hat V }$.

\medskip

\noindent
{\bf Theorem: }
The existence of   a non-trivial second order symmetry operator  implies that $M$ admits a Killing tensor $e^{ab}\not=\la\eta^{ab}$.


%
%
%
%
%
%
%
%
%
%

\section{Non-trivial second order symmetry operators}

A non-trivial second order symmetry operator has the form:
\begin{equation}\label{opeq}
\begin{cases}
\E^{ab}= e^{ab}\one  \\
\F^{a}:= (\ze^{a} +\na_b e^{ab})\one +  \al  \ga^a +\(\ffrac[1/3]\ep_{cb}\na^c e^{ab} \)\ga\\
\G:= (g'  +  i\al  V)\one + \(   ie\_a{}^{b} \na_b V  - {i\eta e^{cb} \na_b\hat V\ep_{ac}}\) \ga^a +\\
\qquad+ \(\ffrac[1/4]\(\na^c \ze^a  -2iq F \ep\^c{}_b e^{ab}   \)\ep_{ca} +  { i\al \hat V}\) \ga\\
\end{cases}
\qquad
\begin{aligned}
& \na^{(c} e^{ab)}=0\\
&\na_a \al=\om_a\\
& \na^{(c} \ze^{a)} =\ffrac[1/2] \La^{ca}\\
& \na_a g'= \La_a
 \end{aligned}
\end{equation}
where we set
\begin{equation}
\begin{cases}
\La^{ca}=  4iq F \ep\^{(c}{}_b e^{a)b} \quad\(\then {\La=0}\)\\
\La_c=  i q F\ep_{ca} \ze^a -\ffrac[1/4]\na_a\(Re^a\_c\)  +   {e\_c{}^{d} \na_d \(V^2 +\eta\hat V^2\)}
\\
 \om_c= 2i  e\_c{}^{d} \na_d V\\
\end{cases}
\end{equation}
and where  the following integrability conditions are satisfied
\begin{equation}\label{iceq}
\begin{cases}
\ep^{bc }\na_{b}\om_{c}=0
	\qquad\then
	\ep^{ac}\na_{c}( e\_a{}^{b} \na_b V)=0\\
\ep^{dc} \na_{d} \La_{c}=0
	\qquad\then
	{
		  i q \ze^a \na_{a}  F =
		2 \ep^{dc}   e\_c{}^{e} \na_e V \na_{d}  V
		-\ffrac[1/4]\ep^{dc} \na_{d}  \na_a\(Re^a\_c\) 
		+ \eta \ep^{dc} \na_{d} \( e\_c{}^b\na_b(\hat V)^2\)
		}
\\
 \ze^a\na_a R =\na_c \(\na^b \La^{da} \)  \ep^{c}\_d\ep_{ab}  \\
\ze^{a}\na_a V =-{\eta\(\ffrac[2/3] \na_a e^{bc}\na_c\hat V \ep\^a{}_b +e^{bc}\na_{ac}\hat V\ep\^a{}_b\)}= -\eta(\epsilon^a_{.b}\nabla_a(e^{bc}\nabla_c\hat V)-\frac 13\epsilon^a_{.b}\nabla_ae^{bc}\nabla_c\hat V)\\
{\ze^a\na_a\hat V}=\ffrac[2/3]\ep_{cb}\na^c e^{ab}  \na_a V+\ep\^c{}_{b}\ e^{ab} \na_{ca} V= \epsilon_{cb}\nabla^c(e^{ab}\nabla_a V)-\frac 13\epsilon_{cb}\nabla^ce^{ab}\nabla_a V =  -\frac 13\epsilon_{cb}\nabla^ce^{ab}\nabla_a V\\
 \end{cases}
\end{equation}

We may rewrite the right-hand set of equations (\ref{opeq}) in configuration coordinates as
\begin{eqnarray}
\nabla^{(\mu}e^{\beta\gamma)}&=&0,\label{ea1}\\
\nabla_\mu \alpha&=&2ie^\beta_\mu\nabla_\beta V,\label{ea2}\\
\nabla^{(\mu}\zeta^{\beta)}&=&2iqF^{(\mu}{ }_{\gamma} e^{\beta)\gamma},\label{ea3}\\
\nabla_\mu g'&=&iqF_{\mu \beta}\zeta^\beta+e^\beta_\mu\nabla_\beta (V^2+\eta \hat V^2)-\frac 14\nabla_\beta(Re^\beta_\mu),\label{ea4}.
\end{eqnarray} 

Some of these equations coincide with the necessary and sufficient conditions for the existence of a constant of motion $K$ quadratic in the momenta of the gauge invariant Hamiltonian 
\begin{equation}\label{Ham}
H=\frac 12 g^{\mu \nu}(p_\mu-iqA_\mu)(p_\nu-iqA_\nu)+U.
\end{equation}
If we assume that $K$ has the gauge invariant form
\begin{equation}\label{Ka}
K=\frac 12 k^{\mu \nu}(p_\mu-iqA_\mu)(p_\nu-iqA_\nu)+B^\mu(p_\mu-iqA_\mu)+ W,
\end{equation}
we obtain
\begin{prop}
$\{H,K\}=0$ is equivalent to 
\begin{eqnarray}
\label{61} \na_{(\mu}k_{\nu \sigma)}&=&0\\
\label{62}\na_{(\mu}B_{\nu)}&=&2iqF_{(\mu}{}^{\sigma}k_{\nu)\sigma}\\
\label{63}\na_\mu W&=&iqF_{\mu \sigma}B^\sigma-2k_\mu{}^{\sigma}\na _\sigma U\\
\label{64}B^\mu \na_\mu U&=&0.
\end{eqnarray}
\end{prop}

A similar form of the first integral is given by Carter \cite{Ca}, who considered  the Hamiltonian for the charged particle orbit when  $U=0$ in (\ref{Ham}). 

We observe that if we set  $k^{\mu \nu}=e^{\mu \nu}$ and $B^\mu=\zeta^\mu$  in (\ref{Ham}) and (\ref{Ka}) then (\ref{61}) and (\ref{62}) agree with (\ref{ea1}) and (\ref{ea3}). On the other hand, if $F_{\mu\nu}$ is equal to zero, $B^\mu=\zeta^\mu$ and $U=\frac \alpha{2i}$ then again (\ref{61}) corresponds to (\ref{ea1}), (\ref{62}) to (\ref{ea3}) and (\ref{63}) to (\ref{ea2}).

An important issue, not strictly related with separation of variables, is to know if a second-order symmetry operator is or is not reducible to the first-order ones, i.e. a linear combinations with constant coefficients of products of the latter. The reducibility happens for example for the Dirac equation on the Minkowski four-dimensional space without external fields, where all second-order symmetry operators of the Dirac equation are reducible \cite{KMW}. An obvious necessary  condition for the reducibility is the existence of non-trivial first-order symmetry operators of the Dirac equation. They are characterized by  the existence of a Killing vector $\xi$ such that (\ref{iceq1}) hold and are of the form (\ref{fokv}).
We know that the non-trivial second-order symmetry operators of the Dirac equation have the pure second-order term
$$
\mathbb E=e^{ab}\nabla_{ab},
$$
where $e^{ab}$ are the components of a Killing tensor. The pure second-order term arising from the symmetrized product of two first-order symmetry operators is
$$
\frac 12 (\xi_r^a\xi^b_s+\xi_s^a\xi_r^b)\nabla_{ab},
$$
where $\xi_r$ and $\xi_s$ are  Killing vectors. Therefore, in order to be reducible, the components of a second-order symmetry operator must be the components of a reducible Killing two-tensor. Consequently we have 

\begin{prop} A  second-order symmetry operator is reducible to a linear combination of products of first-order symmetry operators only if there exist Killing vectors $\xi$ such that
$$
e^{ab}=\sum _{r,s}(\xi^a_r\xi^b_s+\xi^a_s\xi^b_r),
$$
with the additional conditions (\ref{iceq1}), in particular, only if the killing tensor $e$ itself is  reducible. 
\end{prop} 

We recall that in constant-curvature manifolds all Killing tensors are reducible. In other manifolds, for example the skew-ellipsoid, there might exist non-reducible  Killing two-tensors.

\section{Non-trivial second-order symmetry operators in Liouville coordinates}

Our aim is now to obtain the coefficients of the non-trivial second-order symmetry operator (\ref{SO}) on spin manifolds which admit solutions of the Killing tensor equation  (\ref{ea1}).  It is well known that in two-dimension there exists a system of canonical coordinates, called  Liouville coordinates in which the non-zero covariant components of the metric tensor $g_{\mu\nu}$ and the Killing tensor $e_{\mu\nu}$ have the form

$$
g_{11}=A(x)+B(y),\quad g_{00}=\eta g_{11}, 
$$

$$
e_{00}=-g_{00}B(y),\quad e_{11}=g_{11}A(x),
$$
where $A$ and $B$ are arbitrary smooth functions. Liouville coordinates are separable coordinates for the Hamilton-Jacobi, Helmholtz and  Schr\"odinger equations. We now proceed to solve the equations (\ref{iceq})-(\ref{ea4}) for the remaining coefficient functions $\alpha$, $\zeta^b$ and $g'$. The first equation of (\ref{iceq}) is equivalent to $d(e\, dV)=0$ which implies that $V$ has the form  (Appendix D)  
$$
V=g^{00}v_0(x)+g^{11}v_1(y),
$$

where $v_0(x)$ and $v_1(y)$ are arbitrary smooth functions. Thus, the solution of (\ref{ea2}) is
$$
\alpha=2iq\left(e^{00}v_0(x)+e^{11}v_1(y)\right).
$$

We now turn our attention to (\ref{ea3}), where we set $A=0$ anticipating the separation results exposed in Section VII. By solving the PDE system (\ref{ea3}) we obtain the following  solution for $\zeta^1$,
$$
\zeta ^1=-\frac{2f'_1(x)}{\sqrt{B(y)}},
$$  
where $f_1(x)$ is an arbitrary function of integration. Substituting  this solution in (\ref{ea3}) and integrating  
the resulting sistem yields the solution for $\zeta^0$
$$
\zeta^0=f_2(y)-\frac 12 f_1(x)\left(B^{-1/2}\right)',
$$
where $f_2(y)$ is another arbitrary function and we use the integrability condition
$$
qF_{01}=iB^{-5/2}\left(B^2f_1''-\frac 34\eta \left(B^{5/2}f_2'-f_1(B')^2+\frac 23 Bf_1B''\right)\right).
$$
From the third equation of (\ref{iceq}) we obtain $f_1=c_1$. By substituting this value into the fifth of (\ref{iceq}) we get
$$
f_2=-\frac 1{2\eta}B^{5/2}B'(2\eta c_1+\sqrt{\eta B}). 
$$
Therefore, the expression of $F_{01}$ becomes
$$
qF_{01}=\frac i{4\sqrt{\eta}}\left(\frac {B''B-B'}{B^2}\right)=\frac {iB}{4\sqrt{\eta}}R.
$$

By taking in account these results, the second and the fourth equations of (\ref{iceq}) gives the partial derivatives of $\hat V$, then, from the integrability condition we obtain $v_0=0$ and, by integrating,
$$
\hat V=\frac{\sqrt{B^9\eta^5-c_2}}{\sqrt{B}}.
$$  
Similarly, from (\ref{ea4}) we can now integrate  $g'$ as
$$
g'=c_3-\frac 18 \left(\frac {B'}B\right)^2.
$$
We observe that, after all the  assumptions made above, $\alpha=\zeta^1=0$ and $V=g^{11}v_1(y)$. All the conditions of existence for non-trivial second-order symmetry operators are now fulfilled. However, we remark that the symmetry operator obtained in this way is not the most general one,  for example, other solutions can be obtained by imposing conditions on $B$. Moreover, we see in Section VII how to build symmetry operators with zero force field $F$ and unrestricted function $B(y)$ associated with separation of variables. 

\section{Separation of variables}

Following \cite{M,S} we assume that a $m$-component spinor $\psi$ depending on $n$ variables $q^i$ is, locally, multiplicatively separated if 
$$
\psi=\prod_{i=1}^{n}{\phi_i(q^i)}\xi
$$
where $\phi_i(q^i)$ are $n$ diagonal invertible matrices of order $m\times m$ and $\xi$ a constant $m$ spinor. The separation is complete, or regular, if the spinor $\psi$ depends on $nm$ constants $c_i^j$ such that 
$$
det \frac{\partial (\phi^{-1} \partial_i\psi_j)}{\partial c_i^j}\neq 0,
$$
where $\phi=\prod_{i=1}^{n}{\phi_i(q^i)}$. 

This definition is mutuated from the  definition of complete separability for Hamilton-Jacobi and Schr\"odinger equations and essentialy means that a biunivocal correspondence exists between the family of the $\psi$, parametrized by $(c_i^j)$, and the parameters themselves. 

In our case, $n=m=2$ and the independent parameters must be four. We assume $q^0=x,q^1=y$ (from now on our coordinates indices run from $0$ to $1$) and we write a separated spinor as
$$
\psi=\left( \begin{matrix} a_1(x)  b_1(y) \cr a_2(x)b_2(y) \end{matrix} \right).
$$
The completeness condition becomes then 
\begin{equation}
\label{compl}
det \left(\frac {\partial(d\,\ln(a_j)/dx)}{\partial c_i^j}, \frac{\partial(d\,\ln(b_j)/dy)}{\partial c_i^j}\right) \neq 0.
\end{equation}

In \cite{FMRS} and \cite{CFMR} it has been proved that second-order symmetry operators of the Dirac equation in two dimensional Riemannian and pseudo-Riemannian manifolds are strictly associated with second-order Killing tensors. Characteristic Killing tensors (they are KTs with all distinct eigenvalues and orthogonally integrable eigenforms) are associated with orthogonal separable coordinates for the Helmholtz equation. It is known that orthogonal separability of Dirac equation is possible only in Helmholtz-separable orthogonal coordinate systems. Given a second-order symmetry operator,  is then natural to work in the separable coordinates determined  by the Killing tensor associated to the symmetry operator, provided it is a characteristic one. Any symmetric Killing two-tensor is characteristic in Riemannian manifolds of dimension two, provided the Killing tensor is not proportional to the metric tensor. In two-dimensional Lorentzian manifolds there exist Killing 2-tensors not proportional to the metric but not characteristic. We assume here that the Killing tensor associated with the symmetry operator of the Dirac equation is characteristic. We remark that the requirement of reality often made about the eigenvalues and eigenvectors of the Killing tensor is not assumed here. In \cite{DR} and \cite{CFMR} it is shown that the separation of variables in the complex case can be handled essentially in the same way as in the real case. In dimension two, if the Killing tensor is characteristic and if the Robertson condition is satisfied (i.e. the Ricci tensor is diagonalizable simultaneously with $K$ or, equivalently,  the Killing tensor and the Ricci tensor  commute as endomorphisms on the tangent bundle) then the Laplace-Beltrami operator is completely separable in those orthogonal coordinate systems that diagonalize $K$ and the differential operator $\nabla_\mu(K^{\mu \nu}\nabla_\nu)$ commutes with it (see \cite{BCR0,BCR01} and references therein for separation of Helmholtz and Schr\"odinger equations on $n$-dimensional Riemannian or pseudo-Riemannian manifolds). When the separation of variables is associated with first-order symmetry operators \cite{M} we can again relate the separation of variables to second-order symmetry operators, since the square of a first-order symmetry operator is a second-order one. It follows that separation of variables of the Dirac equation, at least in dimension two, can always be understood by means of second-order symmetry operators, as is the case for the separation of the Helmholtz or Schr\"odinger equation.

We remark that in Liouville manifolds the Ricci tensor is always diagonalized in the separable coordinates, therefore the Robertson condition always holds. Hence, 

\prop The existence of a  characteristic Killing 2-tensor $K$ on a Liouville manifold, implies the complete (or regular) separation of the Helmholtz equation in the coordinates associated with $K$. 
\rm

A detailed discussion of complete, regular, non-regular and constrained separation is given in \cite{M, S, BCR0, CR}. For example, in the case of complete or regular separability, the orthogonal separable coordinates for the geodesic Hamilton-Jacobi or Schr\"odinger equations coincide with the St\"ackel systems, Killing tensors determine the separable coordinates and in $\mathbb E^3$ they are the familiar Cartesian, spherical, cylindrical etc coordinates, i.e. confocal quadrics. When  the separation is no longer complete,  the type of separable orthogonal systems changes considerably. For example, in $\mathbb E^3$ with  energy fixed equal to zero, the geodesic Schr\"odinger equation becomes the  Laplace equation, conformal Killing tensors determine the separable coordinates and these  include toroidal, six-spheres, prolate spheroidal  coordinates etc., i.e.  confocal cyclids, that are fourth-degree surfaces \cite{CCM}.

In the following we choose the Dirac representation of the Clifford algebra, namely
$$
\gamma^0=\left(\begin{matrix} 1 & 0 \cr 0 & -1 \end{matrix} \right),\quad \gamma^1=\left(\begin{matrix} 0 & - k \cr k & 0 \end{matrix} \right),\quad 
\gamma=\gamma_0\gamma_1=\left(\begin{matrix} 0 & - \eta k \cr -\eta k & 0 \end{matrix} \right).
$$
The most general covariant potential that can be  present in  the Dirac equation is a combination of an electro-magnetic term $\gamma^\mu A_\mu$, a scalar term $V\mathbb I$ and a pseudoscalar term $\hat V \gamma$, see for example \cite{Th} and references therein, responsible for  nuclear and other type of fermion interactions.
By using the same notation of \cite{CFMR}, the Dirac equation becomes, where we use the form  $(\z^\mu_a)$ of the spin-frame components,
\begin{eqnarray}\label{DE}
\left[\left( \begin{matrix} i \z^0_0 & -ik\z^0_1 \cr ik\z^0_1 & -i \z^0_0\end{matrix} \right) \partial_x +
\left( \begin{matrix} i \z^1_0 & -ik\z^1_1 \cr ik\z^1_1 & i  \z^1_0\end{matrix} \right) \partial_y +
\tilde C\right]\psi =\mu \psi ,
\end{eqnarray}
where $k=i$ for Riemannian and $k=1$ for Lorentzian metrics, i.e. $k=\sqrt{-\eta}$, and
\begin{equation}\label{de4}
\tilde C=\frac i2 \epsilon ^{ab}\z^\mu_a\Gamma^{01}_\mu\gamma_b-q\z^\mu_aA_\mu\gamma^a -V \mathbb I - \hat V \gamma,
\end{equation}

In \cite{MR, FMRS,CFMR}  the multiplicative separation scheme D5 for the general system of first-order partial differential equations of eigenvalue type in two dimensions
\begin{equation}\label{ss}
\mathbb D_5=\left (\begin{matrix} 0 & X_2(x) \cr X_3(x) &0\end{matrix} \right) \partial_x+  \left (\begin{matrix} Y_1(y) & 0 \cr 0 &Y_4(y)\end{matrix} \right) \partial_y+\left (\begin{matrix} C_1(y) & C_2(x) \cr C_3(x) &C_4(y)\end{matrix} \right),
\end{equation}
has been identified as the only separation scheme for the Dirac equation in dimension two uniquely associated with the existence of non-reducible second-order symmetry operators. More precisely,
the Dirac equation in dimension two $\mathbb D$ is multiplicatively separable in coordinates $(x,y)$ iff there exist non-null functions $R_i$ such that
\begin{equation}\label{ssbis}
\mathbb D=\left( \begin{matrix} R_1 & 0 \cr 0 & R_2 \end{matrix} \right) \mathbb D_5 .
\end{equation}
If we assume that the two-spinor solution of the Dirac equation is multiplicatively separated, 
$$
\psi=\left(\begin{matrix}  a_1(x)  b_1(y)\cr a_2(x)b_2(y)\end{matrix}\right),
$$
then the separated equations are
\begin{eqnarray}\label{sep}
\begin{cases}
(X_3\partial_x+C_3)  a_1=\nu_2a_2\\
(X_2\partial_x+C_2)a_2=\nu_1  a_1\\
(Y_1\partial_y+C_1-\mu)  b_1=-\nu_1b_2\\
(Y_4\partial_y+C_4-\mu)b_2=-\nu_2  b_1,
\end{cases}
\end{eqnarray}
From \cite{MR} we know that the symmetry operator $\mathbb K$ associated with the separation scheme D5 is given by the decoupling equations
\begin{eqnarray}\label{decoup0}
\begin{cases}
(X_2\partial_x+C_2)(X_3\partial_x+C_3)  a_1=\nu   a_1,\\
(X_3\partial_x+C_3)(X_2\partial_x+C_2)a_2=\nu a_2,
\end{cases}
\end{eqnarray}
where  $\nu =\nu_1\nu_2$, the product of the separation constants,  is the eigenvalue of the symmetry operator. The remaining decoupling relations
\begin{eqnarray}\label{decoup2}
\begin{cases}
(Y_4\partial_y+C_4-\mu)(Y_1\partial_y+C_1-\mu)  b_1=\nu   b_1\\
(Y_1\partial_y+C_1-\mu)(Y_4\partial_y+C_4-\mu)b_2=\nu b_2,
\end{cases}
\end{eqnarray}
do not define symmetry operators. Since we are interested in the spinor's components $\psi_1, \psi_2$ and not on the single factors $a_i$, $b_i$,  the only conditions to be satisfied by the solutions of the decoupled equations are not the separated equations but
\begin{eqnarray}\label{decoup3}
\begin{cases}
\nu   a_1  b_1=-(X_2 a'_2+C_2a_2)(Y_4 b'_2+C_4b_2-\mu b_2),\\
\nu a_2b_2=-(X_3 a'_1+C_3  a_1)(Y_1 b'_1+C_1  b_1-\mu   b_1),
\end{cases}
\end{eqnarray}
obtained by multiplying together suitable pairs of the separated equations. It is then evident that the  spinor, solution of the Dirac equation, will depend on the parameters $\mu$ and $\nu$, that are the eigenvalues of the Dirac operator and of its symmetry operator, and not on the $\nu_i$ introduced in the separation procedure, where the primes denote derivatives with respect to the arguments.

We observe that the four decoupled equations (\ref{decoup0}) and (\ref{decoup2}) in the four unknown $(a_i)$, $(b_i)$ require, being all second-order differential equations in normal form, a total of eight arbitrary constants in order to be completely integrable (complete or regular separation). Moreover, the two equations (\ref{decoup3}) are, due to the separation of variables, equivalent to four first-order equations in $(a_i)$, $(b_i)$, so that four of the constants in the solutions are determined by them. The remaining parameters are   $\mu$, $\nu$ and two integration constants.

As in \cite{MR}, we do not consider here  the case $\mu = 0$ in full generality. This means that, when $\mu=0$, the conditions of separability given here are sufficient but not necessary, as it is the case for the Hamilton-Jacobi or Schr\"odinger equations with zero energy. From \cite{MR} we have  $R_1(y)$, $R_2(y)$, that we can assume real without restrictions (the dual possibility $R_1(x)$, $R_2(x)$ is equivalent after the exchange $x\leftrightarrow y$). By comparing  (\ref{DE}) and  (\ref{ss})  it follows that 

$$
\z^0_0=\z^1_1=0, \quad \z^0_1=\frac {i}k R_1X_2, \quad \z^1_0=R_1\bar Y_1,
$$
being as in \cite{MR}  $Y_j=i\bar Y_j$ where the $\bar Y_j$ are real functions.
Since the $(\z^\mu_a)$ must be real, we remark that $k=i$ implies $X_2$ real, and $k=1$ implies $X_2$ pure imaginary.
Moreover
$$
R_2=R_1, \quad \bar Y_4=-\bar Y_1,\quad X_3=-X_2,
$$

and
\begin{equation}\label{gg}
g_{00}= \eta(\frac ik R_1 X_2)^{-2}, \quad g_{11}=(R_1\bar Y_1)^{-2}.
\end{equation}

We remark that, up to now, the introduction of external fields does not interfere with the results of \cite{MR} and \cite{CFMR}. The term  term $\tilde C=R_1C$ of the Dirac operator (\ref{de4}) becomes
\begin{eqnarray}\label{tt}
\tilde C=\frac 14\left( \begin{matrix} 2k\z^\mu_1\Gamma^{01}_\mu & 2i\z^\mu_0\Gamma^{01}_\mu \cr -2i\z^\mu_0\Gamma^{01}_\mu & -2k\z^\mu_1\Gamma^{01}_\mu \end{matrix}\right) -
qR_1\left( \begin{matrix} \bar Y_1A_1 & -iX_2 A_0 \cr iX_2 A_0 & -\bar Y_1 A_1 \end{matrix} \right)-
\left( \begin{matrix} V & -\eta k\hat V \cr -\eta k\hat V & V \end{matrix} \right),
\end{eqnarray}


We assume that the mass term $mc^2$, where $c$ is the speed of the light and $m$ is the mass of the particle, is absorbed by the arbitrary constant in the potentials.
We know from \cite{MR} and \cite{CFMR} that separation of the Dirac equation in dimension two is possible only if one of the coordinates is, up to rescaling, geodesically ignorable ($q^i$ is geodesically ignorable iff  $\partial_ig_{jk}=0$ $\forall j,k$) and it is easy to check that the introduction of the external fields is immaterial in this respect. Then, from (\ref{gg}) we have that it is not restrictive to assume $X_2=-i k$, in order to have real spin frame components for any possible value of $k$, so that the metric  depends on $y$ only. Hence, essentially two coordinate systems are associated with this type of separation, the Liouville coordinates and the "polar" coordinates discussed below. 

%

\subsection*{Liouville coordinates}
If we require  that  the metric is in Liouville form  $g_{11}=A(x)+B(y)$, $g_{00}=\eta g_{11}$,   it follows that $A=0$ and  we can set without restrictions  
$$
R_1=\beta(y)^{-1}, \quad X_2=-i k, \quad \bar Y_1=1,
$$ 
so that $B(y)=\beta(y)^{2}$.
Consequently, the Dirac operator becomes
$$
\mathbb D=\frac {ik}\beta\left( \begin{matrix} 0 & -1\cr 1 & 0 \end{matrix}\right)\partial_x+\frac {i}\beta\left( \begin{matrix} 1 & 0\cr 0 & -1 \end{matrix}\right)\partial_y+\left( \begin{matrix} \frac {i\beta '}{2\beta^2}+\frac q\beta A_1-V & -\frac{kq}\beta A_0+ \eta k\hat V \cr \frac{kq}\beta A_0+\eta k\hat V & -\frac {i\beta '}{2\beta^2}-\frac q\beta A_1-V \end{matrix}\right)
$$
and, consequently, we have

\begin{eqnarray}\label{nescsep}
\begin{cases}
qA_0= \frac 1{2k}(C_3(x)-C_2(x)),\\
qA_1= \frac 12 \left(C_1(y)-C_4(y)-i\frac {\beta'}\beta\right),\\ 
V=-\frac 1{2\beta}(C_1(y)+C_4(y))\\
\hat V=\frac 1{2k\eta \beta}(C_2(x)+C_3(x)).
\end{cases}
\end{eqnarray}
From these relations, since $g_{11}=\beta^2$ and in these coordinates $g^{ii}=g_{ii}^{-1}$, we have (see Appendix D)

\begin{prop}\label{exactL} In Liouville coordinates  the vector potential $A_\mu$  separable in the scheme D5  is necessarily exact and the  force field $F_{\mu\nu}=\partial_\mu A_\nu-\partial_\nu A_\mu$ is equal to zero.
\end{prop}

Moreover, 

\begin{prop}\label{sf}
In Liouville coordinates, the scalar and pseudoscalar potentials are compatible with separation of variables in the scheme D5  only if  $V^2$ and $\hat V^2$ are St\"ackel multipliers, that is only if 
$$
d(e\; d(V^2))=0, \quad d(e\;d(\hat V^2))=0.
$$ 
\end{prop}
 
The computation of (\ref{decoup0}) provides the operator
\begin{eqnarray}\label{decoupaL}
\eta \left[\left(-\partial ^2_{xx}+2iq A_0\partial_x+iq \partial_xA_0+q^2A_0^2 -\beta^2  \hat V^2\right)\mathbb I +i\eta \beta \partial_x\hat V \left(\begin{matrix} 1 & 0 \cr 0 & -1 \end{matrix} \right) \right] \psi=\nu\psi,
\end{eqnarray}
and the remaining decoupling relations  (\ref{decoup2}) become
\begin{eqnarray}\label{decoupbL}
\left[\left( \partial^2_{yy} + \left(\frac {\beta'}\beta - 2iq A_1\right)\partial_y+\frac {\beta ''}{2\beta}-\frac 14 \left(\frac {\beta '}{\beta}\right)^2-q^2A^2_1-i\frac {\beta'}\beta qA_1-\right.\right.\nonumber\\
\left. \left.-iq\partial_yA_1+\beta^2V^2+2\mu \beta V +\mu^2\right)\mathbb I+i\partial_y(\beta V)\left(\begin{matrix} 1 & 0 \cr 0 & -1 \end{matrix} \right)\right]\psi=\nu\psi.
\end{eqnarray}
It is possible to check that the separated decoupling operator given by (\ref{decoupaL}) commutes with the Dirac operator \cite{MR}. Therefore, we have

\begin{prop} In Liouville coordinates, the vector, scalar and pseudoscalar potentials are compatible with separation of variables in the scheme D5 associated with a symmetry operator if and only if they are of the form (\ref{nescsep}). 
\end{prop}

We remark that the first-order terms in (\ref{decoupaL}) and (\ref{decoupbL}) disappear if  
\begin{equation}\label{vpcond}
qA_0=0, \quad qA_1=-\frac {i\beta'}{2\beta}.
\end{equation}

The exactness of $(A_\mu)$ assures that such a term can always be introduced without affecting the physics of the system, apart from some effect on the phase of the spinors (Aharonov-Bohm effect) and corresponds to the gauge invariance discussed in Section II.  Applications of the  freedom of choice of a phase factor for the spinor   are made for example in \cite{Co} and \cite{Sh1} in order to simplify the Dirac equation in curvilinear coordinates.

With this choice of $(A_\mu)$,  for any $\beta(y)$ the decoupling relations (\ref{decoupaL}) and (\ref{decoupbL}) give     respectively
\begin{eqnarray}\label{deceq}
\begin{cases}
-\eta \left( a''_1(x) +\beta (\beta \hat V^2-i\eta \partial_x \hat V)\right)  a_1=\nu   a_1(x),\\
-\eta \left( a''_2(x) +\beta (\beta \hat V^2+i\eta \partial_x \hat V)\right)a_2=\nu a_2(x),\\
  b_1''(y)+\left(i\partial_y(\beta V)+(\beta V+\mu)^2\right)  b_1(y)=\nu   b_1(y),\\
b_2''(y)+\left(-i\partial_y(\beta V)+(\beta V+\mu)^2\right)b_2(y)=\nu b_2(y).
\end{cases}
\end{eqnarray}

We now substitute the (\ref{nescsep}), with $g_{00}=\eta \beta^2$, $g_{11}=\beta^2$, $g_{01}=g_{10}=0$, into the  equations (\ref{SOSOP}) determining the  covariant second-order symmetry operator obtained in Section V. After some computations we obtain

\begin{prop}\label{cov} The  second-order symmetry operator (\ref{decoupaL}) associated with the separable Liouville coordinates is determined by the following conditions 
\begin{enumerate}
\item $e$ is the canonical Killing tensor associated with the Liouville coordinates: $e_{00}=-\eta \beta ^4$, $e_{10}=e_{01}=e_{11}=0$.
\item $\alpha$ is zero,
\item $\alpha^\mu$ is the zero vector,
\item $\zeta$ is the zero vector,
\item the function $g'$ is given, up to additive constants, by 
$$ g'=\frac 14 \left( (C_2(x)+C_3(x))^2+\left(\frac {\beta '}{\beta}\right)^2\right).$$
\end{enumerate}
\end{prop}

Remarkably, for $\nu=0$ the equations (\ref{deceq}) are in the form $z''-(w^2+w')z=0$, with $w=\pm i\eta \beta \hat V$ for the $a_i$ and $w=\pm (i\beta V-i\mu)$ for the $b_i$. The general solution of these equations is in this case \cite{PZ}
$$
z(r)=c_1z^0+c_2z^0\int{\frac {dr}{(z^0)^2}},\quad z^0=e^{\int{w(r)\, dr}}.
$$

In general, for $V=\hat V=0$ the decoupling equations of above can be easily integrated, giving, after the imposition of (\ref{sep}) and (\ref{decoup3})
\begin{eqnarray}\label{ex1}
\begin{cases}
\psi_1=\left(c_1 e^{ \frac {\sqrt{\nu}}{\hat k}x}+c_2 e^{ -\frac {\sqrt{\nu}}{\hat k}x}\right)\left(d_1 \sin \sqrt{\mu^2-\nu}y+d_2\cos \sqrt{\mu^2-\nu}y\right),\\
\psi_2=\left(c_3 e^{ \frac {\sqrt{\nu}}{\hat k}x}+c_4 e^{ -\frac {\sqrt{\nu}}{\hat k}x}\right)\left(d_3 \sin \sqrt{\mu^2-\nu}y+d_4\cos \sqrt{\mu^2-\nu}y\right),
\end{cases}
\end{eqnarray}
where $c_3=i(\nu)^{-\frac 12}c_1 $, $c_4=-i(\nu)^{-\frac 12}c_2$, $d_3=d_1\mu+id_2\sqrt{\mu^2-\nu}$, $d_4=d_2\mu-id_1\sqrt{\mu^2-\nu}$.

We remark that the Dirac equation in this case is "geodesic", since no external force field is active, even though a non-null vector potential is present. We remark that, even if four constants $c_i$ are given in (\ref{ex1}), what is relevant for the completeness (\ref{compl}) of the solution are the ratios $a'_i/a_i$ and $b'_i/b_i$. Therefore, two parameters disappear and we are left with the ratios, say, $c_1/c_2$, $d_1/d_2$ and the dynamical parameters $\mu$ and $\nu$ only.

An interesting example of Hamilton-Jacobi and Schr\"odinger equations with scalar potentials separable in these coordinates on curved spaces can be found in \cite{Bal1}.  The classical Hamiltonian given there is considered under different quantizations and represents a generalization of the harmonic oscillator to conformally flat $n$-dimensional Riemannian manifolds. We rewrite it for the two-dimensional Riemannian or pseudo-Riemannian case in Liouville coordinates as
$$
H=\frac {e^{-2y}}{2(1+\lambda e^{2y})}(\eta p_x^2+p_y^2)+ \frac {\omega ^2 e^{2y}}{2(1+\lambda e^{2y})},
$$
where $\lambda$ and $\omega $ are parameters. From the expression of the metric tensor and from (\ref{nescsep}) it is evident that in the corresponding Dirac equation, $C_1, C_4$ can be chosen  so that $V$ coincides with the scalar potential of $H$ and $A_1=-\frac{i \beta'}{2\beta}$, while $C_2=C_3=0$ give $A_0=\hat V=0$.

\subsection*{"Polar" coordinates}
It can be useful for computation to choose coordinates such that the metric is in the form
  $g_{00}=\eta B(y)$, $g_{11}=1 $. We call these "polar"  coordinates. They are related to  Liouville coordinates of above by a simple reparametrization. We remark that in this case also the Robertson condition is satisfied. Then, we can choose 
$$
R_1=\beta(y)^{-1}, \quad X_2=-i k, \quad \bar Y_1=\beta(y).
$$
so that $B(y)=\beta(y)^{2}$. Then, the Dirac operator is
$$
\mathbb D=\frac {ik}\beta\left( \begin{matrix} 0 & -1\cr 1 & 0 \end{matrix}\right)\partial_x+\left( \begin{matrix} i & 0\cr 0 & -i \end{matrix}\right)\partial_y+\left( \begin{matrix} \frac {i\beta '}{2\beta}+ q A_1-V & -\frac{kq}\beta A_0+ \eta k\hat V \cr \frac{kq}\beta A_0+\eta k\hat V & -\frac {i\beta '}{2\beta}- q A_1-V \end{matrix}\right).
$$
Therefore, we have

\begin{eqnarray}\label{nescsep1}
\begin{cases}
qA_0= \frac 1{2k}(C_3(x)-C_2(x)),\\
qA_1= \frac 1{2} \left(C_1(y)-C_4(y)-i \beta'\right),\\ 
V=-\frac 1{2\beta}(C_1(y)+C_4(y))\\
\hat V=\frac 1{2k\eta \beta}(C_2(x)+C_3(x)).
\end{cases}
\end{eqnarray}
Hence,
\begin{prop}\label{exactP} In "Polar" coordinates  the vector potential $A_\mu$  separable in the scheme D5  is necessarily exact and the  force field $F_{\mu\nu}=\partial_\mu A_\nu-\partial_\nu A_\mu$ is equal to zero.
\end{prop}

and, as for the Liouville coordinates, we have
\begin{prop}
In "Polar" coordinates, the scalar and pseudoscalar potentials are compatible with separation of variables in the scheme D5  only if  $V^2$ and $\hat V^2$ are St\"ackel multipliers. That is, only if 
$$
d(e\; d(V^2))=0, \quad d(e\;d(\hat V^2))=0.
$$ 
\end{prop}

The computation of (\ref{decoup0}) gives the operator
\begin{eqnarray}\label{decoupaP}
\eta \left[\left(-\partial ^2_{xx}+2iq A_0\partial_x+iq \partial_xA_0+q^2A_0^2 -\beta^2  \hat V^2\right)\mathbb I +i\eta \beta \partial_x\hat V \left(\begin{matrix} 1 & 0 \cr 0 & -1 \end{matrix} \right) \right] \psi=\nu\psi,
\end{eqnarray}
and the remaining decoupling relations  (\ref{decoup2}) become
\begin{eqnarray}\label{decoupbP}
\left[\left( \beta^2\partial^2_{yy} +2\beta (\beta '-  iq A_1)\partial_y+\frac 12\beta \beta ''+\frac 14 (\beta ')^2-q^2A^2_1-iq\beta\partial_yA_1+\right.\right.\nonumber\\
\left. \left.+\beta^2V^2+2\mu \beta V +\mu^2\right)\mathbb I+i\beta \partial_y(\beta V)\left(\begin{matrix} 1 & 0 \cr 0 & -1 \end{matrix} \right)\right]\psi=\nu\psi.
\end{eqnarray}
Again, the (\ref{decoupaP}) determine a symmetry operator.
We observe that the first-order terms in (\ref{decoupaP}) and (\ref{decoupbP}) disappear if  

\begin{equation}\label{vpcond1}
qA_0=0, \quad qA_1=-i\beta'.
\end{equation}
With this choice, the decoupled equations become
\begin{eqnarray}\label{deceq2}
\begin{cases}
-\eta \left( a''_1(x) +\beta (\beta \hat V^2-i\eta \partial_x \hat V)\right)  a_1=\nu   a_1(x),\\
-\eta \left( a''_2(x) +\beta (\beta \hat V^2+i\eta \partial_x \hat V)\right)a_2=\nu a_2(x),\\
\beta^2 b_1''(y)+\left(\left(\frac {\beta'}2\right)^2 +i\beta\partial_y(\beta V)+(\beta V+\mu)^2\right)  b_1(y)=\nu   b_1(y),\\
\beta^2 b_2''(y)+\left(\left(\frac {\beta'}2\right)^2-i\beta \partial_y(\beta V)+(\beta V+\mu)^2\right)b_2(y)=\nu b_2(y).
\end{cases}
\end{eqnarray}

The second-order operator is  determined by the same  decoupling relations  as for the Liouville coordinates.

\begin{prop} In "Polar" coordinates, the vector, scalar and pseudoscalar potentials are compatible with separation of variables in the scheme D5 associated with a symmetry operator if and only if they are of the form (\ref{nescsep1}). 
\end{prop}
The substitution of these conditions into the into the  equations (\ref{SOSOP}) in the case of "polar" coordinates gives
\begin{prop}\label{cov1} The  second-order symmetry operator (\ref{decoupaP}) associated with the separable "Polar" coordinates is determined by the following conditions 
\begin{enumerate}
\item $e$ is the canonical Killing tensor associated with the Polar coordinates (the same as for the Liouville coordinates): $e_{00}=-\eta \beta ^4$, $e_{10}=e_{01}=e_{11}=0$.
\item $\alpha$ is zero,
\item $\alpha^\mu$ is the zero vector,
\item $\zeta$ is the zero vector,
\item the function $g'$ is given, up to additive constants, by 
$$ g'=\frac 14 \left( (C_2(x)+C_3(x))^2+(\beta ')^2\right).$$
\end{enumerate}
\end{prop}

By choosing $\beta=y$, we get $g_{00}=\eta y^2$, $g_{11}=1$,  and we are dealing with the true polar coordinates in Euclidean or Minkowski plane. The scalar potential $V=\frac h y$ determines the Kepler-Coulomb system. In this case, our Dirac equation with (\ref{vpcond1}) and $\hat V=0$ yields the solution
\begin{eqnarray}\label{solP}
\begin{cases}
\psi_1=\left(c_5e^{ \frac {\sqrt{\nu}}{ k}x}+c_6e^{ -\frac {\sqrt{\nu}}{\hat k}x}\right)\left(c_1 y^{\frac 12 +w}+c_2y^{\frac 12-w}\right)\\
\psi_2=\left(ic_5e^{ \frac {\sqrt{\nu}}{ k}x}-ic_6e^{ -\frac {\sqrt{\nu}}{\hat k}x}\right)\left(\frac{h+\mu-iw}{\sqrt{\nu}}c_1y^{\frac 12+w}+\frac{h+\mu+iw}{\sqrt{\nu}}c_2y^{\frac 12-w}\right),
\end{cases}
\end{eqnarray}
where $w=\sqrt{\nu-(h+\mu)^2}$.

By introducing the functions
$$
S_\kappa(z)=\left\{\begin{array}{ll}
\frac{\sin\sqrt{\kappa}z}{\sqrt{\kappa}} & \kappa>0 \\
z & \kappa=0 \\
\frac{\sinh\sqrt{|\kappa|}z}{\sqrt{|\kappa|}} & \kappa<0
\end{array}\right.
\qquad
C_\kappa(z)=\left\{\begin{array}{ll}
\cos\sqrt{\kappa}z & \kappa>0 \\
1 & \kappa=0 \\
\cosh\sqrt{|\kappa|}z & \kappa<0
\end{array}\right., \quad T_\kappa(z)=\frac {S_\kappa(z)}{C_\kappa(z)},
$$
the trigonometric and hyperbolic functions can be treated simultaneously. 
By setting $\beta(y)=S_\kappa(y)$, the  metric $g_{ii}$ defines the sphere $\mathbb S^2$, the Euclidean plane $\mathbb E^2$, the hyperbolic plane $\mathbb H^2$ for $\kappa=1,0,-1$ respectively and $\eta=1$, while the choice $\eta=-1$ gives the anti-de Sitter, the Minkowski  and the de Sitter two-dimensional spaces for  $\kappa=1,0,-1$ respectively. 
An example of a classical  Hamiltonian on curved manifolds with associate Dirac equation which is separable in "polar" coordinates is 
$$
H=\frac 12\left(\eta p_x^2+\frac 1{S_\kappa ^2(y)}p_y^2\right)+\alpha_1 T_\kappa ^2(y)+\alpha_2 T^{-1}_\kappa (y),
$$
 representing (for $\eta=1$) a curved Higgs  oscillator ($\alpha_2 =0$) or a curved Kepler-Coulomb system ($\alpha_1 =0$) on $\mathbb S^2$ or $\mathbb H^2$ according to  $\kappa =1$ or $\kappa =-1$ respectively (see for example \cite{Bal2}). Here, the coordinates $(x,y)$ are geodesic polar coordinates. Obviously, a rescaling of the coordinates can put the metric tensor in Liouville form, so that  $H$ is also separable in Liouville coordinates, yielding the same expression (\ref{ex1}) for the solutions with $\hat V=V=0$ and $(A_\mu)$ chosen  according to (\ref{vpcond}). If we choose $\beta=S_\kappa(y)$, then $T_\kappa ^{-1}=\beta'/\beta$. For $V=T^{-1}_\kappa(y)$,  the decoupling equations (\ref{deceq2}) can be solved for $b_1$, $b_2$ in term of hypergeometric functions, while $a_1$, $a_2$ have  the same form as in (\ref{solP}).

\section{Conclusion}

We give several necessary conditions for the separation in Liouville coordinates  of the Dirac equation associated with second-order symmetry operators in an invariant form: the existence of a Killing two-tensor $e$ with a Killing eigenvector, an exact vector potential, scalar and pseudoscalar potentials $V$ and $\hat V$ satisfying $d(e\;dV)=d(e\;dV^2)=0$, $d(e\;d\hat V^2)=0$. The correspondence between the second-order symmetry operator and the decoupling operator generated by the separation of variables is made explicit. 

\section*{Appendix A}
Let us here collect identities which are used to expand the symmetry condition $[\K,\D]=0$ to obtain the equations (\ref{Cond}).
\begin{equation}
\begin{cases}
[D_\mu, D_\nu ]\psi= \ffrac[1/4] R^{ab}{}_{\mu\nu} \ga_{ab}\psi -iq F_{\mu\nu} \psi \\
[D_\mu, D_\nu ]D_\al \psi= \ffrac[1/4] R^{ab}{}_{\mu\nu} \ga_{ab}D_\al \psi -iq F_{\mu\nu} D_\al\psi - R^\la{}_{\al\mu\nu} D_\la \psi \\
\end{cases}
\label{CommNabla}
\end{equation}
These are both proven by expanding the covariant derivatives by using (\ref{CovDer}) and recollecting the curvature tensors.

We also have some Lemmas to expand the iterated covariant derivatives in the basis of symmetrized covariant derivatives.
\begin{equation}
\begin{cases}
D_{a}D_b \psi= D_{ab}\psi +\ffrac[1/8] R^{ef}{}_{ab} \ga_{ef}\psi -\ffrac[iq/2]F_{ab}\psi\\
D_{ab}D_c \psi= D_{abc}\psi 
+\ffrac[1/6] \na_{(a} R^{ef}{}_{b)c} \ga_{ef} \psi
+\ffrac[1/4] \ga_{ef} D_{(a}\psi R^{ef}{}_{b)c}
-\ffrac[1/3]  R^{e}{}_{(ab)c}D_e\psi+\\
\qquad\qquad\quad
-iq F_{c(a} D_{b)}\psi
-\ffrac[2iq/3]\na_{(a} F_{b)c}\psi
\\
D_c D_{ab} \psi= D_{abc}\psi
-\ffrac[1/12] \na_{(a} R^{ef}{}_{b)c} \ga_{ef} \psi
-\ffrac[1/4] \ga_{ef} D_{(a}\psi R^{ef}{}_{b)c}
+\ffrac[2/3]  R^{e}{}_{(ab)c}D_e\psi+\\
\qquad\qquad\quad
-iq F_{c(a} D_{b)}\psi
+\ffrac[iq/3]\na_{(a} F_{b)c}\psi
\\
\end{cases}
\end{equation}

Here $D_{abc}\psi=D_{(a}D_bD_{c)}\psi$ denotes the symmetrized triple covariant derivative.
These are proven by using commutators of covariant derivatives written in terms of the curvature (See (\ref{CommNabla})).

\section*{Appendix B}

We shall here collect useful formulae to manipulate products of Dirac matrices in dimension $m=2$ in any signature $\eta$.
Hereafter $\eta$, by  abuse of language, denotes also the determinant of $\eta_{ab}$.

\begin{equation}
(\ga)^2= -\eta \one
\qquad
\ga_c\ga= -\ga \ga_c=\eta \ep_{ca} \ga^a
\qquad
\ga_a\ga_b= \eta_{ab}\one + \ep_{ab} \ga
\end{equation}
\begin{equation}
[\ga_{cd}, \ga_a]= 4\eta_{a[d} \ga_{c]}
\end{equation}
\begin{equation}
[\ga_{cd}, \ga_{ab}]= 2\eta_{cb} \ga_{ad}+2\eta_{ac} \ga_{db}-2\eta_{db} \ga_{ac}-2\eta_{da} \ga_{cb}
\end{equation}

\section*{Appendix C}
Le us review some identities about Killing vectors and tensors in dimension 2.
Among other things they are used to simplify equations (\ref{eq31}).

Let us start by considering a vector $\ze^a$ which satisfies an equation of the form
\begin{equation}
\na^{(a} \ze^{b)}= \ffrac[1/2] \La^{\al\be}
\end{equation}
for some symmetric tensor $\La^{\al\be}$. 
For $\La^{\al\be}=0$ this reproduce the Killing equation.

One can easily prove that 
\begin{equation}
2\na^a\na^{b} \ze^{c}= R(\eta^{ac}\ze^b- \eta^{ab}\ze^c) + \na^a \La^{bc} -\na^c \La^{ab} + \na^b \La^{ac}
\end{equation}
From this, one obtains by contraction
\begin{equation}
\begin{cases}
2\na_a\na^{a} \ze^{c}= -R\ze^c + 2\na_a \La^{ac} -\na^c \La\_{a}{}^{a} \\
2\na_a\na_{b} \ze^{c}= R\ze_b  + \na_b \La\_a{}^{a}\\
\end{cases}
\end{equation}

Let us now consider a symmetric tensor $e^{ab}$.
By expanding the commutator $\ep_{dc} [\na_a, \na^d] e^{ac}$ one can show that
\begin{equation}
\ep_{dc} \na_a\na^d e^{ac}=\ep_{dc} \na^d\na_a e^{ac}
\end{equation}
Let us now assume that $e^{ab}$ is a Killing tensor. We have the following identities on second derivatives
\begin{equation}
\begin{aligned}
& \ep_{dc} \na_a\na^d e^{ac}=0\\
& \na_b \na^c e^{ab}- \na_b\na^b e^{ac}= 2 \na_b \na^c e^{ab} + \na_b \na^a e^{cb} \\
 \end{aligned}
\end{equation}
This last identity can be split into the symmetric and antisymmetric parts to obtain
\begin{equation}
\begin{aligned}
& \na_b \na^{(c} e^{a)b}- \na_b\na^b e^{ac}= 3\na^{(c} \na_b  e^{a)b} + 3R(e^{ac} -\ffrac[1/2] e^b{}\_b \eta^{ac}) \\
& \na_b \na^{[c} e^{a]b}= \na^{[c} \na_b  e^{a]b}  \\
 \end{aligned}
 \label{L1}
\end{equation}

\section*{Appendix D}
We recall here the essentials of the geometric theory of the separation of variables for Hamilton-Jacobi, Helmholtz and Schr\"odinger equations. The results exposed here can be found in \cite{Ben, BCR,BCR0} and references therein. Given a natural Hamiltonian $H=\frac 12 g^{ij}p_ip_j+V(q^i)$ on a Riemannian or pseudo-Riemannian space of metric $(g^{ij})$ and orthogonal coordinates $(q^i)$, the Hamilton-Jacobi equation associated with $H$ has an additively separable solution in $(q^i)$ if and only if the Levi-Civita equations hold
$$
p_ip_j\left(\frac 12 S_{ij}(g^{kk})p_k^2+S_{ij}(V)\right)=0,\quad i\neq j,
$$
where the so-called St\"ackel operator $S_{ij}$ is defined by 
$$
S_{ij}(A)=g^{ii}g^{jj}\partial_{ij}A-g^{ii}\partial_ig^{jj}\partial_jA-g^{jj}\partial_jg^{ii}\partial_iA=0, \quad i \neq j,
$$
and no sum is made over repeated indices. This is equivalent to 
$$
S_{ij}(g^{kk})=0,\quad S_{ij}(V)=0,
$$
the last equation, sometimes called "Bertrand-Darboux equation", is equivalent to $V=g^{ii}\phi_i(q^i)$, where each $\phi_i(q^i)$ is any function of $q^i$ only, $V$ is then called a "St\"ackel multiplier".
It is possible to prove \cite{Ben} that Levi-Civita equations are the integrability conditions of the Eisenhart equations
$$
g^{jj}\partial_i\rho_j=(\rho_i-\rho_j)\partial_i \ln g^{jj},\quad i,j\, n.s.
$$
where $(\rho_i)$ are the pointwise distinct eigenvalues of a symmetric Killing two-tensor $e$, whose eigenvectors are orthogonal to the coordinate hypersurfaces, and therefore  $e^{ij}=\delta ^{ij}\rho_ig^{ii}$. This establishes a correspondence between Killing two-tensors with pointwise distinct eigenvalues and normal (i.e. surface-forming) eigenvectors and orthogonal coordinates separating additively the integrals of the Hamilton Jacobi equation of $H$. 
A symmetric Killing 2-tensor $e$ with distinct eigenvalues and normal eigenvectors (or eigenforms) characterizes completely an orthogonal separable coordinate system and is called "characteristic". This is true even when some of the eigenvalues of $e$ are complex (and, since $e$ is real, exist in complex conjugate pairs) \cite{DR}. The introduction of a characteristic Killing tensor allows one to write the equations $S_{ij}(V)=0$ in the equivalent invariant form
$$
d(e\;dV)=0,
$$
where $d$ is the exterior derivative.

The same characterization holds for the orthogonal coordinates  separating multiplicatively the Helmholtz or Schr\"odinger equations determined by the Laplacian of $(g^{ij})$ and by the scalar potential $V$ \cite{BCR0}. In this case, the separation of these equation is possible if and only if the corresponding Hamilton-Jacobi equation is additively separable and   the additional Robertson condition holds. It is possible to prove that the Robertson condition is equivalent to the diagonalization of the Ricci tensor in the orthogonal coordinates separating  the geodesic Hamilton-Jacobi equation \cite{BCR0}. It is easy to show that in two-dimensional  Liouville coordinates and  in the "polar" coordinates considered in the article the Ricci tensor is always diagonalized, then that the Robertson condition holds.

\section*{Acknowledgements}

The authors wish to thank their reciprocal institutions, the Dipartimento di Matematica, Universit\`a di Torino
and the Department of Applied Mathematics, University of Waterloo for hospitality during which parts of this
paper were written.  The research was supported in part by a Discovery Grant
from the Natural Sciences and Engineering Reasearch Council of Canada and by a Senior Visiting Professorships of the Gruppo Nazionale di Fisica Matematica - GNFM-INdAM (RGM).

\end{document}